\documentclass[%
 aps,
 prb,
 amsmath,amssymb,
 floatfix,
preprint,%
showpacs,
superscriptaddress
]{revtex4-1}

\usepackage{graphicx}
\usepackage{dcolumn}
\usepackage{bm}
\usepackage{hyperref}

\usepackage {epstopdf}

\begin{document}

\title{Impact of silicon doping on low frequency charge noise and conductance drift in GaAs/AlGaAs nanostructures}
\author{S. Fallahi}
\email{sfallahi@purdue.edu}
\affiliation{Department of Physics and Astronomy, Purdue University, West Lafayette, IN}
\affiliation{Birck Nanotechnology Center, Purdue University, West Lafayette, IN}

\author{J. R. Nakamura}
\affiliation{Department of Physics and Astronomy, Purdue University, West Lafayette, IN}
\affiliation{Birck Nanotechnology Center, Purdue University, West Lafayette, IN}

\author{G. C. Gardner}
\affiliation{Birck Nanotechnology Center, Purdue University, West Lafayette, IN}
\affiliation{School of Materials Engineering, Purdue University, West Lafayette, IN}
\affiliation{Station Q Purdue, Purdue University, West Lafayette, IN}

\author{M. M. Yannell}
\affiliation{School of Electrical and Computer Engineering, Purdue University, West Lafayette, IN}

\author{M. J. Manfra}
\email{mmanfra@purdue.edu}
\affiliation{Department of Physics and Astronomy, Purdue University, West Lafayette, IN}
\affiliation{Birck Nanotechnology Center, Purdue University, West Lafayette, IN}
\affiliation{School of Materials Engineering, Purdue University, West Lafayette, IN}
\affiliation{Station Q Purdue, Purdue University, West Lafayette, IN}
\affiliation{School of Electrical and Computer Engineering, Purdue University, West Lafayette, IN}

\date{\today}

\begin{abstract}
We present measurements of low frequency charge noise and conductance drift in modulation doped GaAs/AlGaAs heterostructures grown by molecular beam epitaxy in which the silicon doping density has been varied from $2.4\times 10^{18} cm^{-3}$ (critically doped) to $6.0\times 10^{18} cm^{-3}$ (overdoped). Quantum point contacts were used to detect charge fluctuations. A clear reduction of both short time scale telegraphic noise and long time scale conductance drift with decreased doping density was observed. These measurements indicate that the {\it neutral} doping region plays a significant role in charge noise and conductance drift. 
\end{abstract}

\maketitle

Semiconductor nanostructures such as quantum dots (QDs) and quantum point contacts (QPCs) are essential building blocks of mesoscopic devices used to realize solid state qubits \cite{Martins:2016,Nichol:2017,Avinun:2005,Petta:2005,Li:2004}. Molecular beam epitaxy (MBE) growth is a mature technology for growing extremely pure GaAs/AlGaAs two-dimensional electron gases (2DEGs) with minimal defects. Nevertheless, devices with metallic Schottky gates fabricated on GaAs/AlGaAs heterostructures often suffer from 1/f-like noise and random telegraph noise (RTN) which degrade device performance and may preclude stable device operation. Low frequency noise is believed to arise from time-dependent fluctuations in the occupation of charge trapping sites in the vicinity of the nanostructure that results in fluctuations of the local electrostatic potential. The etiology and dynamics of these charge trapping sites are not fully understood.

Prior work on charge noise in laterally gated nanostructures suggests that the charge noise is mainly due to either (1) current leakage through the Schottky barrier \cite{Cobden:1992,Pioro:2005,Buizert:2008,Liang:2011} or (2) electron hopping within the doping layer \cite{Timp:1990,Kurdak:1997}. Techniques such as bias cooling \cite{Pioro:2005} and use of global top gates \cite{Buizert:2008} suppress charge noise by shifting the operation point to less negative voltage. These studies suggest leakage from gates plays a significant role in generation of low frequency noise. However, these techniques complicate device operation and do not address the root cause of the phenomena. 

A second problem frequently afflicting gated GaAs heterostructures is drift in device conductance over long time scales upon initial cool-down to cryogenic temperatures. This drift makes operating mesoscopic devices difficult, requiring frequent retuning. Conductance drift has not been studied as extensively as short time scale charge noise. In this work, we systematically investigate the relationship between modulation doping density and both low frequency charge noise and drift in device conductance. Both conductance drift and charge noise are shown to depend strongly on the density of silicon doping. We discuss the underlying mechanisms; our results will inform future designs of quiet semiconductor platforms for quantum information research.

\begin{figure}[h]
  \includegraphics[width=0.5\textwidth]{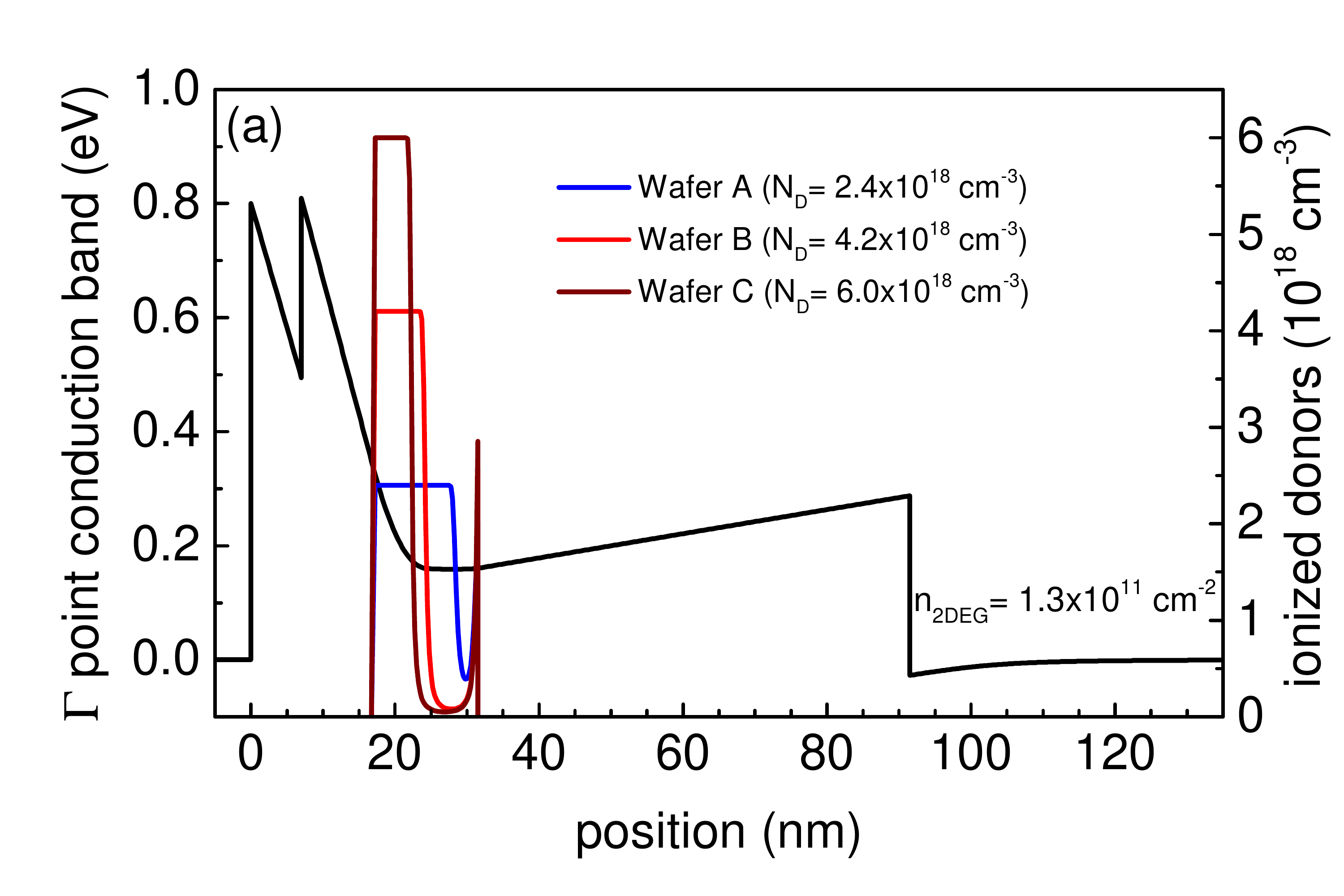}
  \caption{Simulated heterostructure band diagram for uniformly doped single interface heterojuntion with varying doping density. Colored solid lines are the ionized donor profile for each doping concentration. }
  \label{Fig1}
\end{figure}

In order to study the effect of doping concentration on low frequency charge noise, modulation doped GaAs/Al$_{0.36}$Ga$_{0.64}$As heterostructures with three different silicon doping densities N$_{D}$, $2.4\times10^{18}$cm$^{-3}$(Wafer A) , $4.2\times10^{18}$cm$^{-3}$(Wafer B) and $6.0\times10^{18}$cm$^{-3}$(Wafer C) were investigated. These uniformly doped single interface heterostructures were grown by MBE  with a 60nm AlGaAs spacer between the 2DEG and the doping region (14.5nm thick) and total 90nm 2DEG depth measured from the top surface. An overview of sample parameters is given in Table 1. In Fig. \ref{Fig1} (a) we show the conduction band profile for wafer B simulated using the Nextnano software package \cite{Birner:2007}. We also show the ionized donor density for wafers A, B, and C. 

\begin{table*}[h]
\caption{Characteristics of studied wafers including Si doping concentration N$_{D}$, doping width W, 2DEG density n, mobility $\mu$, operating gate voltage V$_{g}$ and number of QPCs measured for each wafer.}
\label{Table 1}
\begin{ruledtabular}
\begin{tabular}{ccccccc}
\hline
 Wafer Name & N$_{D}$ (cm$^{-3}$) & W (nm) &  n (cm$^{-2}$) &  $\mu$ (cm$^{2}$/Vs) & V$_{g}$ (V) & $\#$ of QPCs measured   \\  \hline \hline
 Wafer A & 2.4$\times$10$^{18}$  &  14.5 & 1.1$\times$10$^{11}$ & 2.7$\times$10$^{6}$ & -0.5 & 6  \\ \hline
 Wafer B & 4.2$\times$10$^{18}$  &  14.5 & 1.3$\times$10$^{11}$ & 5.1$\times$10$^{6}$ & -0.6 & 3 \\ \hline
 Wafer C & 6.0$\times$10$^{18}$  &  14.5 & 1.4$\times$10$^{11}$ & 4.7$\times$10$^{6}$ & -0.6 & 6 \\  \hline
\end{tabular}
\end{ruledtabular}
\end{table*}

Three distinct regions exist in the doping layer for each wafer: (1) a positively charged region closer to the cap layer that compensates surface states and produces a Schottky barrier eV$_{b}$$\sim 0.8$ eV at the surface; (2) a neutral region in the middle of the doping layer, where the Fermi level is located at an energy E$_{D}$$\sim 150$ meV \cite{Schubert:1984,Chadi:1989} below the conduction band edge; and (3) a thin ($<$1nm) positively ionized layer from which electrons have been transferred to the 2DEG.
Microscopically, the neutral region is believed to be composed of positively and negatively charged Si donors with almost the same concentration. According to negative-U model proposed independently by Chadi and Chang \cite{Chadi:1989}, and Khachaturyan, Weber, and Kaminska \cite{Khachaturyan:1989}, the substitutional Si donor in Al$_x$Ga$_{1-x}$As (for x$>$0.22) occupying a Ga site has two possible electronic states: 1) a shallow donor level E$_{d}$ with no lattice relaxation and 2) a deep and localized donor level E$_{DX}$ with large lattice relaxation which binds two electrons. Based on the negative U-model, half of the donors in the neutral region are positively charged (ionized) shallow d$^{+}$ and the remaining half are negatively charged DX$^{-}$ states.

Importantly for our experiments, the doping width is kept constant at 14.5nm for the three wafers A, B and C; only the silicon doping density is varied. As charge transfer to the 2DEG is determined by the constant conduction band offset and setback, an increase in doping density does not significantly change the 2DEG density or the charge transferred to the surface. Rather, the width of the neutral region increases, as is seen by comparing the black, red, and pink traces in Fig. \ref{Fig1}. Wafer A is close to critical doping (meaning that nearly all dopants are positively ionized), while Wafer C is significantly overdoped. Due to the presence of DX centers, the electrons in this neutral region can be frozen at low temperatures (below 100K) \cite{Kunzel:1983,Kunzel:1984}; no parallel conduction is observed in magnetotransport measurements (not shown).

\begin{figure}[h]
  \includegraphics[width=0.5\textwidth]{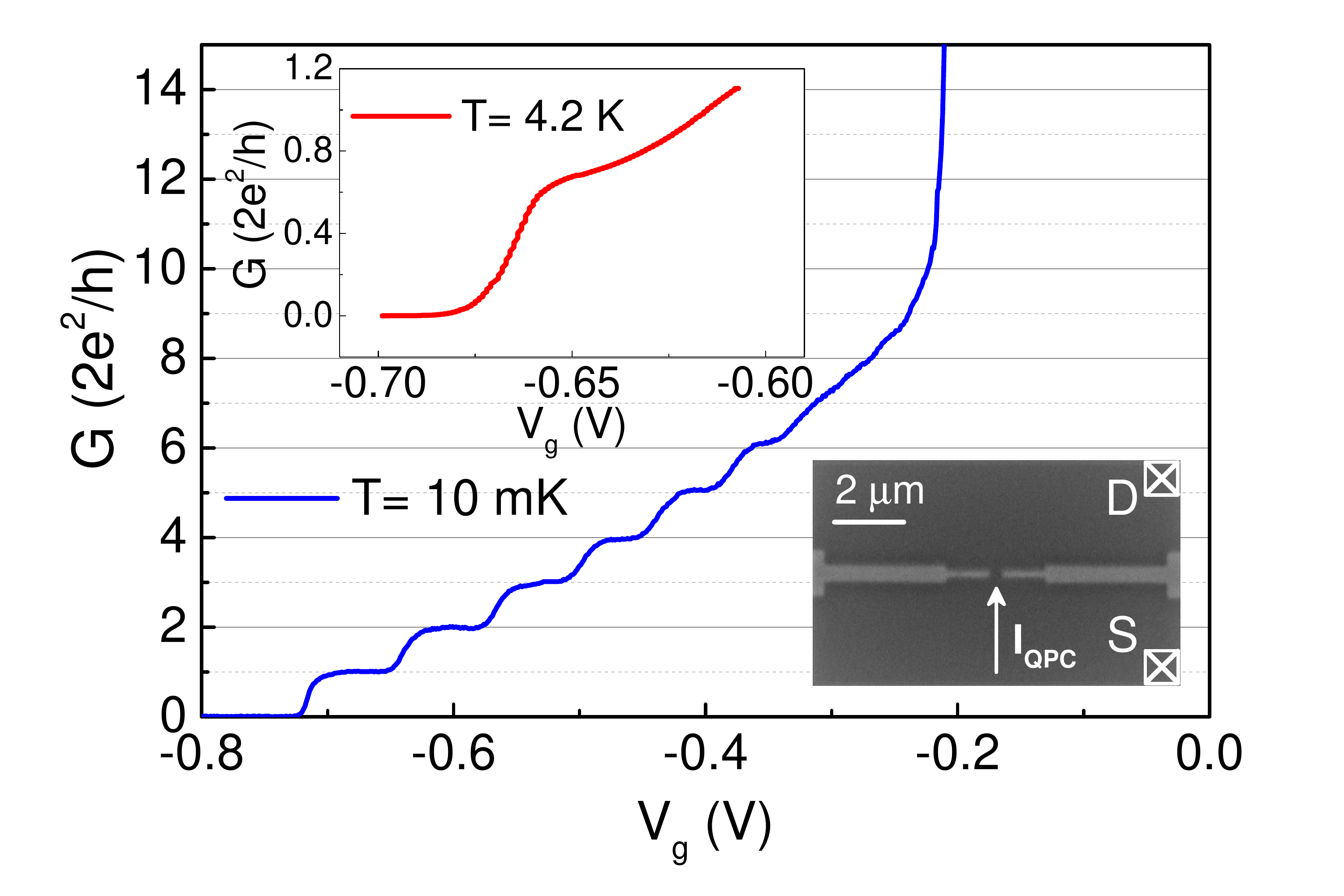}
  \caption{Conductance G of a QPC as a function of gate voltage $V_{g}$ at T=10mK. Top inset: first riser of QPC conducatance at T=4.2K. Bottom inset: SEM image of fabricated QPC.}
  \label{Fig2}
\end{figure}

We utilized QPCs as charge sensors to detect charge noise. QPCs with a nominal width of 300nm were fabricated on all wafers using identical fabrication procedures to compare the level of charge noise for each wafer. An SEM image of a typical QPC is shown in the inset to Fig. \ref{Fig2}. The processing steps are as follows: (1) photolithography of mesa pattern and mesa etch, (2)  photolithography of ohmic contacts; evaporation of Ni/Au/Ge metal contacts and annealing (3) Electron beam lithography and evaporation of  QPC gates (4) photolithography and evaporation of bonding pads to wire bond devices to a chip carrier for measurement.

Figure \ref{Fig2} shows a typical conductance plot of a QPC as a function of gate voltage V$_{g}$ taken in a dilution refrigerator with a mixing chamber plate temperature T=10mK; the conductance is quantized in units of $2e^{2}/h$ corresponding to discrete conductance modes of the device. Bias cooling is not employed in any of our experiments. The gate voltage required to deplete the 2DEG beneath the gates is essentially identical for all the studied wafers and equal to -185 mV. The geometric capacitance between the gate and 2DEG is C=$\epsilon_0 \epsilon_r/$d per unit area where d equals the 2DEG depth beneath the top surface. Assuming only coupling between the gate and 2DEG, we calculated the depletion gate voltage V$_{dep}$=$e$n/C=-180mV for n=1.3$\times 10^{11}$ cm$^{-2}$. This nearly perfect agreement implies that charges in the neutral region do not respond to gate voltage and are frozen at low temperature. 

The top inset shows the first riser in QPC conductance at T = 4.2K, where we operated the devices for noise measurements. At T = 4.2K, the QPC still has very high transconductance on the riser of the first quantized conductance plateau, making it very sensitive to the position of individual charges in the vicinity of the device. We used a two-terminal measurement in which a 200 $\mu$V DC voltage bias is applied to the source contact, and drain current is measured with a DL1211 current pre-amplifier; the output is fed to a National Instruments NI-DAQ digitizer.  

\begin{figure}[h]
  \includegraphics[width=0.5\textwidth]{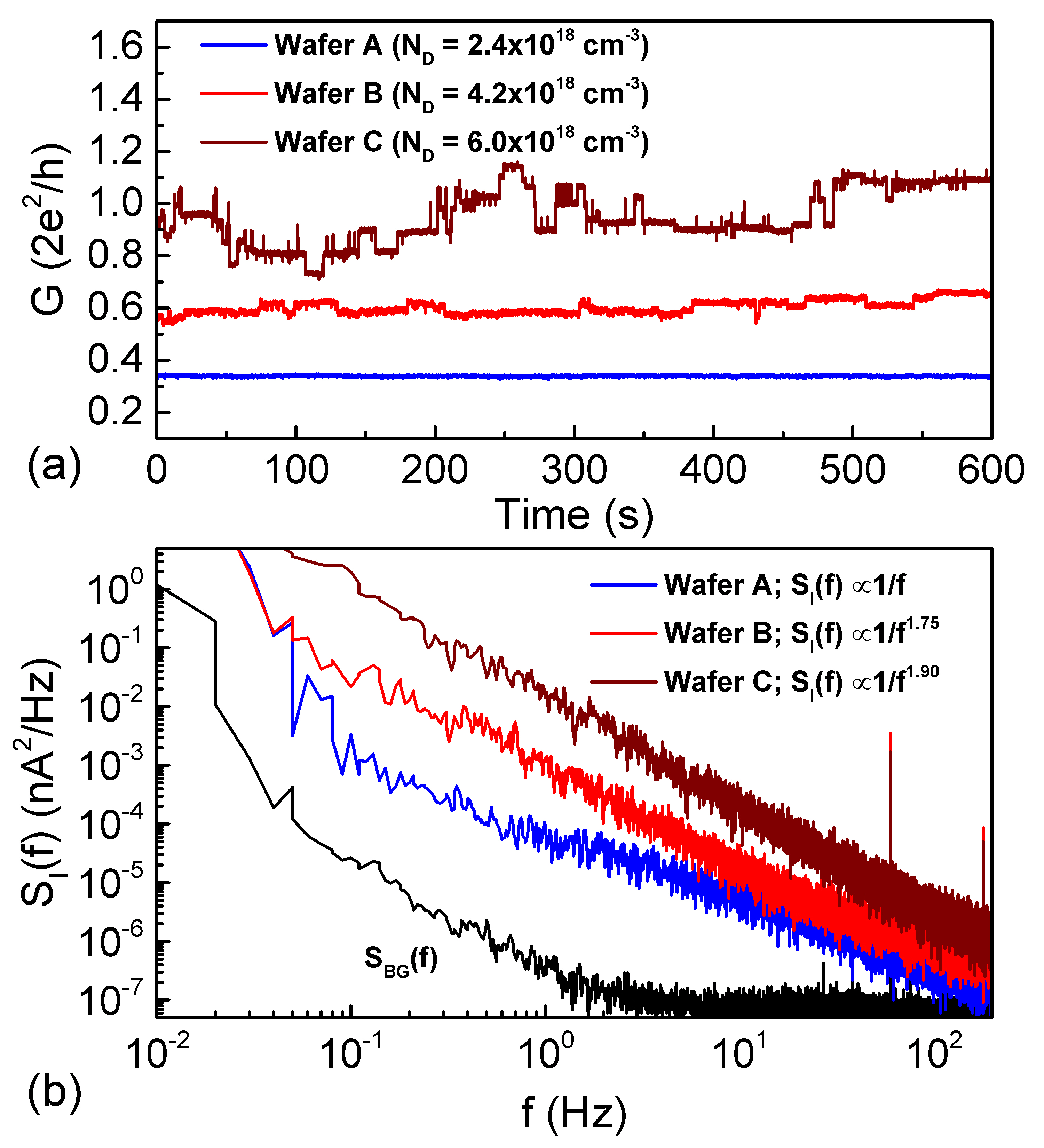}
  \caption{(a) QPC time traces for structures with different doping densities, offset for clarity. (b) noise power spectral density $S_{I}(f)$ obtained from FFT of time traces, experimental background noise $S_{I;BG}(f)$ measured at zero source drain voltage (black trace).}
  \label{Fig3}
\end{figure}

The most striking observation of our study is the dramatic increase in low frequency noise associated with increased doping density as shown in Fig. \ref{Fig3}. Conductance time traces for QPCs sitting at the first riser of conductance are shown in Fig. \ref{Fig3} (a) for QPCs from wafers A, B, and C. Note that the operating gate voltage is nearly indentical in all three cases. The conductance of the QPC on wafer A is nearly constant, indicating that this QPC suffers minimal charge noise. The QPC from wafer B shows increased noise and discrete switching events, while the QPC from wafer C shows significant noise amplitude and severe RTN visible in the raw data. Clearly, the level of charge noise increases as the doping density is increased. The noise power spectral density, obtained from a Fast Fourier Transform (FFT) of the time traces, is shown in  Fig. \ref{Fig3} (b).  For comparison, the noise power spectrum of the measurement circuit with zero source-drain bias applied to the device is also shown. The increase of RTN as doping density is increased is reflected in the frequency dependence of power spectral density, which shifts from 1/f (for the lowest doping density) indicative of a broad ensemble of trapping sites with a homogeneous distribution of switching time scales to Lorentzian dependence 1/f$^2$ (for the highest doping density) indicative of the strong influence proximal two-level traps \cite{McWhorter:1957}. 

We quantify the noise level for each wafer in terms of equivalent gate voltage noise $\Delta V_g$, given in Eq. \ref{Eq1} (this represents the voltage noise level applied on the QPC gates that would produce the same conductance fluctuations as caused by the charge noise) \cite{Jung:2004}. In Eq. \ref{Eq1}, $S_{I}(f)$ is the power spectral density of current fluctuations through the QPC and $S_{I;BG}(f)$ is background noise due to noise in our instruments.

\begin{align}
\Delta V_{g} = \sqrt[]{\int_{0.1}^{100} [S_{I}(f)-S_{I;BG}(f)] df} \bigg / \left(\frac{dI_{QPC}}{dV_{g}}\right)
\label{Eq1}
\end{align}

\begin{figure}[h]
   \includegraphics[width=0.5\textwidth]{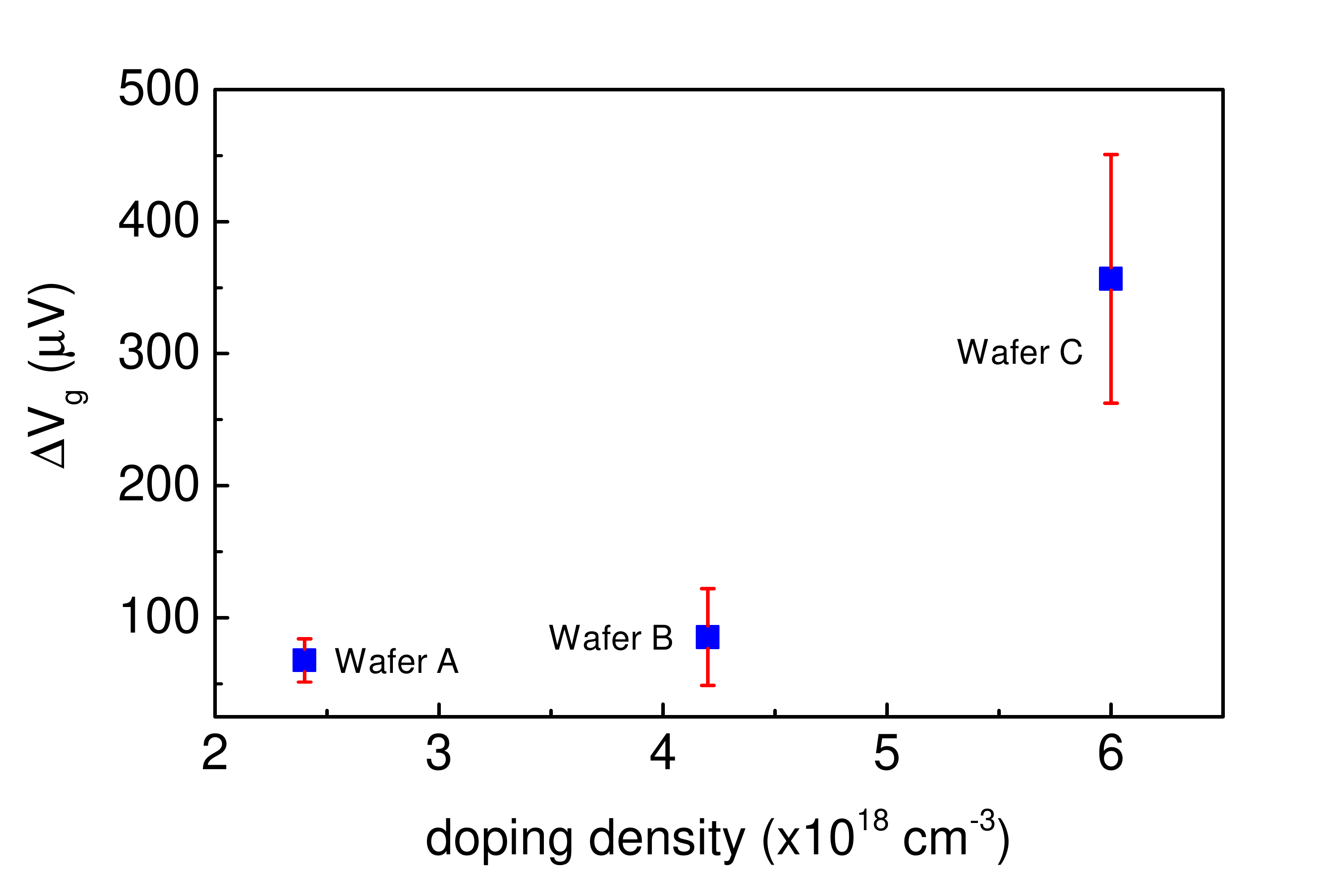}
  \caption{Equivalent gate voltage noise $\Delta V_{g}$ vs. doping density for Wafers A, B and C. Each data point represents the average of multiple different QPCs from each wafer. Six QPCs were measured from Wafers A and C, and three QPCs were measured from Wafer B. Error bars represent the standard error computed from the measurements of different QPCs from each wafer. }
  \label{Fig4}
\end{figure}

Fig. \ref{Fig4} shows the equivalent gate voltage noise vs. doping density for each wafer. Each data point represents the average of multiple different QPCs from each wafer; six QPCs were measured from Wafer A, three were measured from Wafer B, and six were measured from Wafer C. This plot shows a correlation between the noise level and doping density. In particular, the equivalent gate voltage noise is substantially larger for the highest doping density wafer, Wafer C, as is the device to device variation as indicated by the increase of the standard error.

According to the negative-U model, the neutral region of the doping layer is expected to contain ionized shallow donors (d$^+$) that may act as trapping sites and contribute to charge noise. Prevailing theory suggests that electrons tunneling from the Schottky gates are temporarily trapped on these sites and contribute to noise. As our heterostructures are essentially identical apart from the doping density, the operating voltages are nearly identical. This implies that the tunneling matrix element for electrons leaking from the surface gate are the same for all three wafers. Since the noise clearly increases as a function of doping density, we propose that the {\it number} of trapping sites (shallow ionized d$^+$ donors) within the neutral region has a primary impact. The noise level increases due to the increasing width of the neutral region and corresponding increase in available donor states.  We are in essence increasing the final density of states for the tunneling process which leads to enhanced low frequency noise.  While this analysis clearly suggests that heterostructures should be minimally doped to reduce low frequency noise, other considerations including formation of low resistance ohmic contact and production of high mobility 2DEGs make determination of optimal doping a subtle optimization problem.

\begin{figure}[h]
   \includegraphics[width=0.5\textwidth]{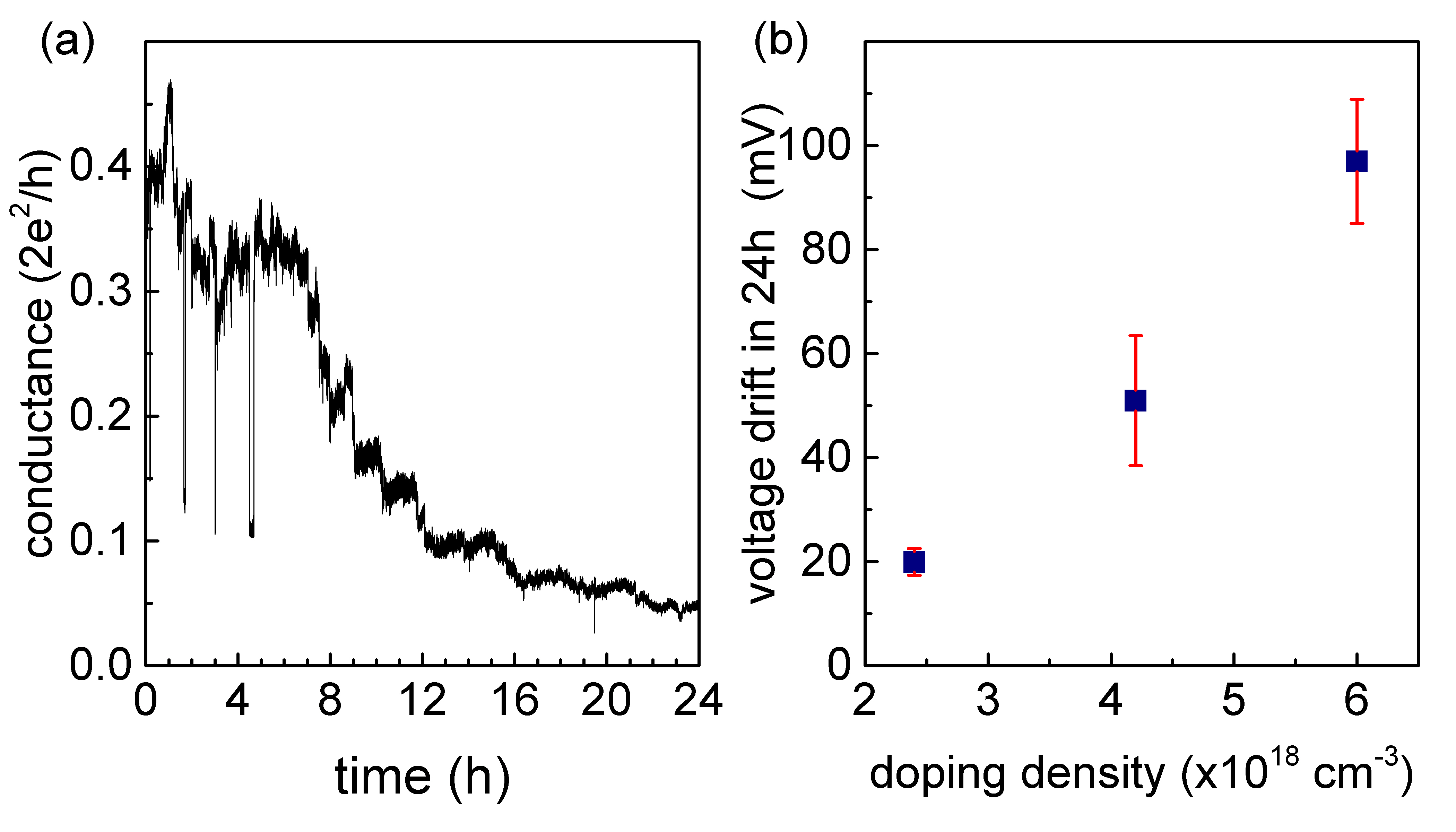}
  \caption{(a) Long time scale conductance drift. (b) Total amount of gate voltage drift within 1st 24 hours of operation of QPCs for wafers A, B and C with different doping densities.}
  \label{Fig5}
\end{figure}

The second phenomenon we investigated is drift in conductance over long time scales at fixed gate voltage. A typical conductance time trace upon initial cool down for a QPC sitting on the first riser of conductance plateau is shown in Fig. \ref{Fig5} (a). Although both negative and positive jumps in conductance occur, the overall trend is that conductance \textit{decreases} over time at fixed gate bias; equivalently, the operating point of the QPC shifts to \textit{less negative} voltage over time. We observed this trend for all QPCs cooled with the gates grounded and then energized at T=4.2K. Typically the largest amount of drift occurs within the first 24 hours after initially biasing the QPC at low temperature, after which the conductance starts to saturate.

We quantify the amount of drift exhibited by each sample as the shift in gate voltage required to operate the QPC on the first conductance riser after 24 hours. This quantity is plotted for each wafer in Fig. \ref{Fig5} (b); the data is from the same QPCs which were used to characterized noise. As with the RTN, it is clear that the level of QPC drift increases with increasing doping density.

Our data suggests that the drift phenomenon may be understood in the following way. Applying negative voltage to the surface gates raises the chemical potential at the gate, $\mu_{gate}$, relative to the chemical potential of the 2DEG, $\mu_{2DEG}$ that is connected to ground. Because the doping layer lies between the gate and the 2DEG, the chemical potential at the doping layer will tend to increase so that it is intermediate between $\mu_{gate}$ and $\mu_{2DEG}$, leading to an increase over time in the average occupation of donor states. Each time an electron tunnels to a donor site near the QPC, the repulsive potential causes a \textit{negative} jump in the conducatance of the QPC. However, because of the substantial tunneling barrier between the surface and the doping layer, the average occupation of donor sites increases slowly; the chemical potential at the doping layer slowly rises as electrons tunnel to the available donor states before saturating at a steady-state value. The dynamics and saturation of this long time scale behavior may also be impacted by the complex electric field configuration in the immediate vicinity of the gate edges where the the electric field has both vertical and horizontal components. Additionally, the fact that the drift occurs over time scales much longer than the RTN suggests that drift may involve deep donor levels with a barrier to electron capture \cite{Chadi:1989}, whereas RTN may primarily involve shallow donor d$^+$ levels. 

\begin{figure}[h]
   \includegraphics[width=0.5\textwidth]{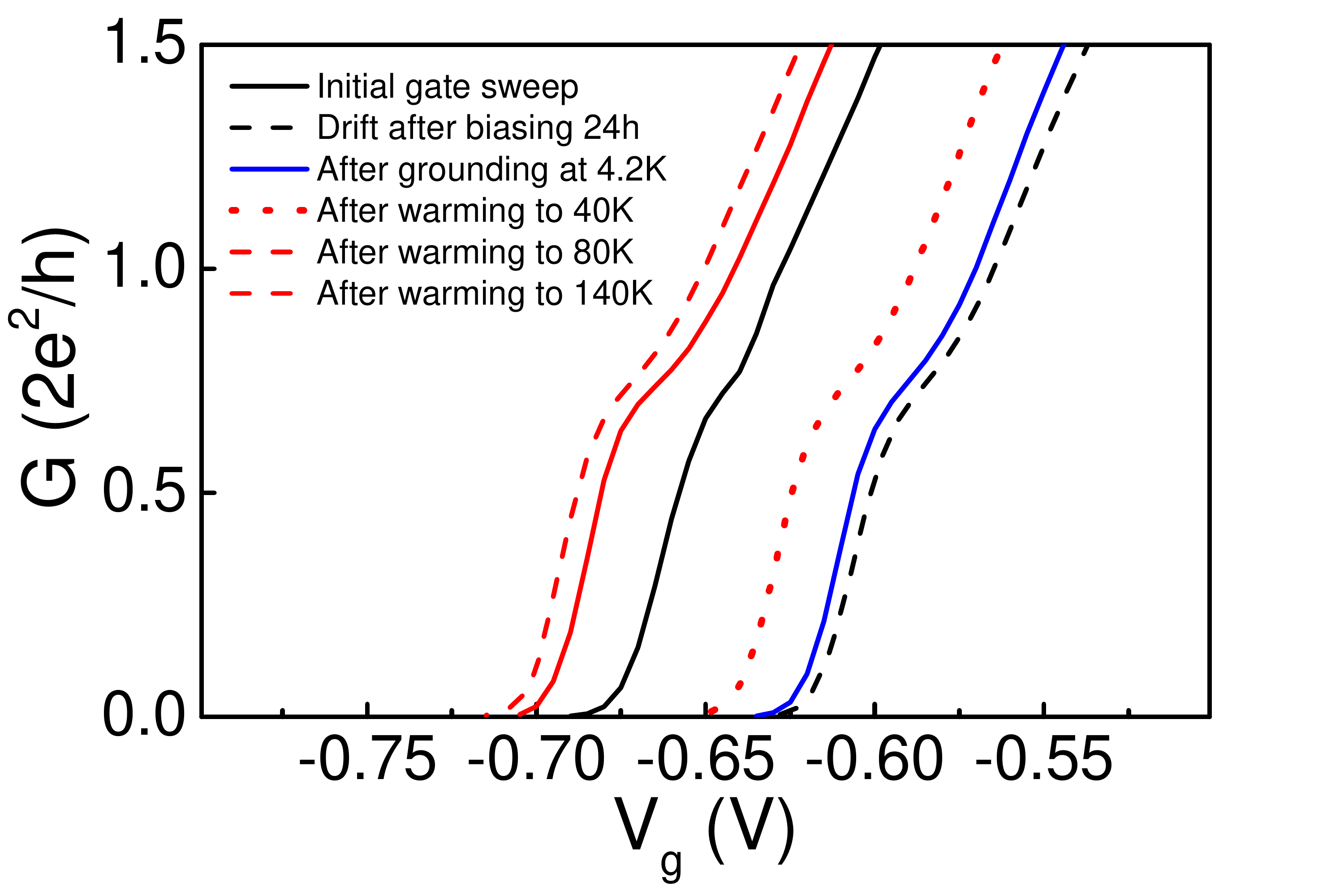}
  \caption{QPC conductance plot vs gate voltage for Wafer B after warming up the QPC to T= 40K, 80K and 140K.}
  \label{Fig6}
\end{figure}

We performed an additional experiment to investigate the temperature stability of the charge accumulation associated with the conductance drift phenomenon. A QPC from wafer B was biased on the riser of the first conductance plateau at T = 4.2K for 24 hours; significant conductance drift occurred during this period, consistent with the trend shown in Fig. \ref{Fig5}. The shift in the conductance vs. gate voltage curve due to drift is shown in Fig. \ref{Fig6} (dashed black line compared to solid black line). The QPC was then swept to zero gate bias, and kept at zero gate bias at T = 4.2K for an additional 24 hour period. Next, the QPC gate bias was again swept to obtain the conductance vs. gate voltage curve (blue line in Fig. \ref{Fig6}). After being kept at zero bias for 24 hours, the conductance vs. gate voltage curve did not return to the original state before the drift occurred, but remained shifted and closely matched the curve \textit{after} the drift occurred. This indicates that at T = 4.2K, the accumulated charge that contributes to the conductance drift is frozen; it does not relax after the gate bias is  removed. Next, we warmed the QPC to a series of increasingly higher temperatures: 40K, 80K, and 140K. The QPC was kept at zero gate bias and held at each temperature for approximately 20 hours; immediately after this period, the QPC was cooled to T = 4.2K and its conductance vs. gate voltage characteristics were measured. After warming to 40K, the conductance vs. gate voltage curve shifted to more negative bias (dotted red line), but did not return all the way to its original state before the drift occurred, indicating that a significant fraction, but not all, of the charge accumulated due to drift remained frozen in place at T=40K. After warming to 80K, the conductance curve (dashed red line) shifted to even more negative bias beyond the initial pre-drift curve. We take this as an indication that the majority of donor states that have trapped electrons in the vicinity of the QPC are now thermally depopulated. Warming to 140K resulted in a slight shift in the conductance vs. gate voltage curve (solid red line). We attribute the small difference in the initial gate sweep at 4.2K and the sweep after warming the sample to T=140K to random rearrangement of donors states as is typically seen in the majority of QPCs upon thermal cycling to room temperature.

The fact that the charge accumulated in the drift process remains frozen at T = 4.2K after the gate bias is removed indicates that the donor state involved in conductance drift has a barrier to emission. The DX$^-$ donor state traps an electron and is known to have a barrier to emission; however, charge in DX$^-$ states remain frozen at temperatures below 100K \cite{Kunzel:1983,Kunzel:1984}. The fact that we observe partial thermal depopulation at 40K and full depopulations at 80K suggests that the state responsible for conductance drift is shallower than the  DX$^-$ state. Evidence for a trap state associated with the Si donors with a smaller barrier to emission than the DX$^-$ state was reported in Ref. \cite{Jia:1989}; it is plausible that these states could be responsible for the drift we observe.   

In conclusion, we unambiguously identified the total number of silicon donors as an important parameter influencing low frequency charge noise and conductance drift in modulation doped GaAs/AlGaAs heterostructures. Our data suggests that electron tunneling to available donor states, especially those in the neutral region, contributes to charge noise and device drift. The comparatively short time scale of the charge noise implies that it primarily involves shallow donor states, while the much longer time scale and the apparent freezing of the charge involved in drift suggests that the drift involves deep donor states. Modulation doped GaAs/AlGaAs heterostructures should be grown close to critical doping (that is, with no unionized donors) to minimize the number of charge trap sites available. Using this guideline, devices with minimal charge noise may be achieved that can serve as a robust, stable platform for spin-qubit based quantum computing.  

\section*{Acknowledgments} 
This work was supported by the U.S. DOE Office of Basic Energy Sciences, Division of Materials Sciences and Engineering Award No. DE-SC0006671.  Additional support from Nokia Bell Laboratories for the MBE facility is gratefully acknowledged. M. Manfra acknowledges many useful conversations with David Reilly concerning charge noise in mesoscopic semiconductor devices.


\begin{thebibliography}{20}%
\makeatletter
\providecommand \@ifxundefined [1]{%
 \@ifx{#1\undefined}
}%
\providecommand \@ifnum [1]{%
 \ifnum #1\expandafter \@firstoftwo
 \else \expandafter \@secondoftwo
 \fi
}%
\providecommand \@ifx [1]{%
 \ifx #1\expandafter \@firstoftwo
 \else \expandafter \@secondoftwo
 \fi
}%
\providecommand \natexlab [1]{#1}%
\providecommand \enquote  [1]{``#1''}%
\providecommand \bibnamefont  [1]{#1}%
\providecommand \bibfnamefont [1]{#1}%
\providecommand \citenamefont [1]{#1}%
\providecommand \href@noop [0]{\@secondoftwo}%
\providecommand \href [0]{\begingroup \@sanitize@url \@href}%
\providecommand \@href[1]{\@@startlink{#1}\@@href}%
\providecommand \@@href[1]{\endgroup#1\@@endlink}%
\providecommand \@sanitize@url [0]{\catcode `\\12\catcode `\$12\catcode
  `\&12\catcode `\#12\catcode `\^12\catcode `\_12\catcode `\%12\relax}%
\providecommand \@@startlink[1]{}%
\providecommand \@@endlink[0]{}%
\providecommand \url  [0]{\begingroup\@sanitize@url \@url }%
\providecommand \@url [1]{\endgroup\@href {#1}{\urlprefix }}%
\providecommand \urlprefix  [0]{URL }%
\providecommand \Eprint [0]{\href }%
\providecommand \doibase [0]{http://dx.doi.org/}%
\providecommand \selectlanguage [0]{\@gobble}%
\providecommand \bibinfo  [0]{\@secondoftwo}%
\providecommand \bibfield  [0]{\@secondoftwo}%
\providecommand \translation [1]{[#1]}%
\providecommand \BibitemOpen [0]{}%
\providecommand \bibitemStop [0]{}%
\providecommand \bibitemNoStop [0]{.\EOS\space}%
\providecommand \EOS [0]{\spacefactor3000\relax}%
\providecommand \BibitemShut  [1]{\csname bibitem#1\endcsname}%
\let\auto@bib@innerbib\@empty
\bibitem [{\citenamefont {Martins}\ \emph {et~al.}(2016)\citenamefont
  {Martins}, \citenamefont {Malinowski}, \citenamefont {Nissen}, \citenamefont
  {Barnes}, \citenamefont {Fallahi}, \citenamefont {Gardner}, \citenamefont
  {Manfra}, \citenamefont {Marcus},\ and\ \citenamefont
  {Kuemmeth}}]{Martins:2016}%
  \BibitemOpen
  \bibfield  {author} {\bibinfo {author} {\bibfnamefont {F.}~\bibnamefont
  {Martins}}, \bibinfo {author} {\bibfnamefont {F.~K.}\ \bibnamefont
  {Malinowski}}, \bibinfo {author} {\bibfnamefont {P.~D.}\ \bibnamefont
  {Nissen}}, \bibinfo {author} {\bibfnamefont {E.}~\bibnamefont {Barnes}},
  \bibinfo {author} {\bibfnamefont {S.}~\bibnamefont {Fallahi}}, \bibinfo
  {author} {\bibfnamefont {G.~C.}\ \bibnamefont {Gardner}}, \bibinfo {author}
  {\bibfnamefont {M.~J.}\ \bibnamefont {Manfra}}, \bibinfo {author}
  {\bibfnamefont {C.~M.}\ \bibnamefont {Marcus}}, \ and\ \bibinfo {author}
  {\bibfnamefont {F.}~\bibnamefont {Kuemmeth}},\ }\href {\doibase
  10.1103/PhysRevLett.116.116801} {\bibfield  {journal} {\bibinfo  {journal}
  {Phys. Rev. Lett.}\ }\textbf {\bibinfo {volume} {116}},\ \bibinfo {pages}
  {116801} (\bibinfo {year} {2016})}\BibitemShut {NoStop}%
\bibitem [{\citenamefont {Nichol}\ \emph {et~al.}(2017)\citenamefont {Nichol},
  \citenamefont {Orona}, \citenamefont {Harvey}, \citenamefont {Fallahi},
  \citenamefont {Gardner}, \citenamefont {Manfra},\ and\ \citenamefont
  {Yacoby}}]{Nichol:2017}%
  \BibitemOpen
  \bibfield  {author} {\bibinfo {author} {\bibfnamefont {J.~M.}\ \bibnamefont
  {Nichol}}, \bibinfo {author} {\bibfnamefont {L.~A.}\ \bibnamefont {Orona}},
  \bibinfo {author} {\bibfnamefont {S.~P.}\ \bibnamefont {Harvey}}, \bibinfo
  {author} {\bibfnamefont {S.}~\bibnamefont {Fallahi}}, \bibinfo {author}
  {\bibfnamefont {G.~C.}\ \bibnamefont {Gardner}}, \bibinfo {author}
  {\bibfnamefont {M.~J.}\ \bibnamefont {Manfra}}, \ and\ \bibinfo {author}
  {\bibfnamefont {A.}~\bibnamefont {Yacoby}},\ }\href {\doibase
  10.1038/s41534-016-0003-1} {\bibfield  {journal} {\bibinfo  {journal} {npj
  Quantum Information}\ }\textbf {\bibinfo {volume} {3}},\ \bibinfo {pages} {3}
  (\bibinfo {year} {2017})}\BibitemShut {NoStop}%
\bibitem [{\citenamefont {Avinun-Kalish}\ \emph {et~al.}(2005)\citenamefont
  {Avinun-Kalish}, \citenamefont {Heiblum}, \citenamefont {Zarchin},
  \citenamefont {Mahalu},\ and\ \citenamefont {Umansky}}]{Avinun:2005}%
  \BibitemOpen
  \bibfield  {author} {\bibinfo {author} {\bibfnamefont {M.}~\bibnamefont
  {Avinun-Kalish}}, \bibinfo {author} {\bibfnamefont {M.}~\bibnamefont
  {Heiblum}}, \bibinfo {author} {\bibfnamefont {O.}~\bibnamefont {Zarchin}},
  \bibinfo {author} {\bibfnamefont {D.}~\bibnamefont {Mahalu}}, \ and\ \bibinfo
  {author} {\bibfnamefont {V.}~\bibnamefont {Umansky}},\ }\href
  {http://dx.doi.org/10.1038/nature03899} {\bibfield  {journal} {\bibinfo
  {journal} {Nature}\ }\textbf {\bibinfo {volume} {436}},\ \bibinfo {pages}
  {529} (\bibinfo {year} {2005})}\BibitemShut {NoStop}%
\bibitem [{\citenamefont {Petta}\ \emph {et~al.}(2005)\citenamefont {Petta},
  \citenamefont {Johnson}, \citenamefont {Taylor}, \citenamefont {Laird},
  \citenamefont {Yacoby}, \citenamefont {Lukin}, \citenamefont {Marcus},
  \citenamefont {Hanson},\ and\ \citenamefont {Gossard}}]{Petta:2005}%
  \BibitemOpen
  \bibfield  {author} {\bibinfo {author} {\bibfnamefont {J.~R.}\ \bibnamefont
  {Petta}}, \bibinfo {author} {\bibfnamefont {A.~C.}\ \bibnamefont {Johnson}},
  \bibinfo {author} {\bibfnamefont {J.~M.}\ \bibnamefont {Taylor}}, \bibinfo
  {author} {\bibfnamefont {E.~A.}\ \bibnamefont {Laird}}, \bibinfo {author}
  {\bibfnamefont {A.}~\bibnamefont {Yacoby}}, \bibinfo {author} {\bibfnamefont
  {M.~D.}\ \bibnamefont {Lukin}}, \bibinfo {author} {\bibfnamefont {C.~M.}\
  \bibnamefont {Marcus}}, \bibinfo {author} {\bibfnamefont {M.~P.}\
  \bibnamefont {Hanson}}, \ and\ \bibinfo {author} {\bibfnamefont {A.~C.}\
  \bibnamefont {Gossard}},\ }\href {\doibase 10.1126/science.1116955}
  {\bibfield  {journal} {\bibinfo  {journal} {Science}\ }\textbf {\bibinfo
  {volume} {309}},\ \bibinfo {pages} {2180} (\bibinfo {year}
  {2005})}\BibitemShut {NoStop}%
\bibitem [{\citenamefont {Li}\ \emph {et~al.}(2004)\citenamefont {Li},
  \citenamefont {Ren}, \citenamefont {Xiong}, \citenamefont {von Molnár},
  \citenamefont {Ohno},\ and\ \citenamefont {Ohno}}]{Li:2004}%
  \BibitemOpen
  \bibfield  {author} {\bibinfo {author} {\bibfnamefont {Y.}~\bibnamefont
  {Li}}, \bibinfo {author} {\bibfnamefont {C.}~\bibnamefont {Ren}}, \bibinfo
  {author} {\bibfnamefont {P.}~\bibnamefont {Xiong}}, \bibinfo {author}
  {\bibfnamefont {S.}~\bibnamefont {von Molnár}}, \bibinfo {author}
  {\bibfnamefont {Y.}~\bibnamefont {Ohno}}, \ and\ \bibinfo {author}
  {\bibfnamefont {H.}~\bibnamefont {Ohno}},\ }\href
  {http://link.aps.org/doi/10.1103/PhysRevLett.93.246602} {\bibfield  {journal}
  {\bibinfo  {journal} {Physical Review Letters}\ }\textbf {\bibinfo {volume}
  {93}},\ \bibinfo {pages} {246602} (\bibinfo {year} {2004})}\BibitemShut
  {NoStop}%
\bibitem [{\citenamefont {Cobden}\ \emph {et~al.}(1992)\citenamefont {Cobden},
  \citenamefont {Savchenko}, \citenamefont {Pepper}, \citenamefont {Patel},
  \citenamefont {Ritchie}, \citenamefont {Frost},\ and\ \citenamefont
  {Jones}}]{Cobden:1992}%
  \BibitemOpen
  \bibfield  {author} {\bibinfo {author} {\bibfnamefont {D.~H.}\ \bibnamefont
  {Cobden}}, \bibinfo {author} {\bibfnamefont {A.}~\bibnamefont {Savchenko}},
  \bibinfo {author} {\bibfnamefont {M.}~\bibnamefont {Pepper}}, \bibinfo
  {author} {\bibfnamefont {N.~K.}\ \bibnamefont {Patel}}, \bibinfo {author}
  {\bibfnamefont {D.~A.}\ \bibnamefont {Ritchie}}, \bibinfo {author}
  {\bibfnamefont {J.~E.~F.}\ \bibnamefont {Frost}}, \ and\ \bibinfo {author}
  {\bibfnamefont {G.~A.~C.}\ \bibnamefont {Jones}},\ }\href
  {http://link.aps.org/doi/10.1103/PhysRevLett.69.502} {\bibfield  {journal}
  {\bibinfo  {journal} {Physical Review Letters}\ }\textbf {\bibinfo {volume}
  {69}},\ \bibinfo {pages} {502} (\bibinfo {year} {1992})}\BibitemShut
  {NoStop}%
\bibitem [{\citenamefont {Pioro-Ladrière}\ \emph {et~al.}(2005)\citenamefont
  {Pioro-Ladrière}, \citenamefont {Davies}, \citenamefont {Long},
  \citenamefont {Sachrajda}, \citenamefont {Gaudreau}, \citenamefont
  {Zawadzki}, \citenamefont {Lapointe}, \citenamefont {Gupta}, \citenamefont
  {Wasilewski},\ and\ \citenamefont {Studenikin}}]{Pioro:2005}%
  \BibitemOpen
  \bibfield  {author} {\bibinfo {author} {\bibfnamefont {M.}~\bibnamefont
  {Pioro-Ladrière}}, \bibinfo {author} {\bibfnamefont {J.~H.}\ \bibnamefont
  {Davies}}, \bibinfo {author} {\bibfnamefont {A.~R.}\ \bibnamefont {Long}},
  \bibinfo {author} {\bibfnamefont {A.~S.}\ \bibnamefont {Sachrajda}}, \bibinfo
  {author} {\bibfnamefont {L.}~\bibnamefont {Gaudreau}}, \bibinfo {author}
  {\bibfnamefont {P.}~\bibnamefont {Zawadzki}}, \bibinfo {author}
  {\bibfnamefont {J.}~\bibnamefont {Lapointe}}, \bibinfo {author}
  {\bibfnamefont {J.}~\bibnamefont {Gupta}}, \bibinfo {author} {\bibfnamefont
  {Z.}~\bibnamefont {Wasilewski}}, \ and\ \bibinfo {author} {\bibfnamefont
  {S.}~\bibnamefont {Studenikin}},\ }\href
  {http://link.aps.org/doi/10.1103/PhysRevB.72.115331} {\bibfield  {journal}
  {\bibinfo  {journal} {Physical Review B}\ }\textbf {\bibinfo {volume} {72}},\
  \bibinfo {pages} {115331} (\bibinfo {year} {2005})}\BibitemShut {NoStop}%
\bibitem [{\citenamefont {Buizert}\ \emph {et~al.}(2008)\citenamefont
  {Buizert}, \citenamefont {Koppens}, \citenamefont {Pioro-Ladrière},
  \citenamefont {Tranitz}, \citenamefont {Vink}, \citenamefont {Tarucha},
  \citenamefont {Wegscheider},\ and\ \citenamefont
  {Vandersypen}}]{Buizert:2008}%
  \BibitemOpen
  \bibfield  {author} {\bibinfo {author} {\bibfnamefont {C.}~\bibnamefont
  {Buizert}}, \bibinfo {author} {\bibfnamefont {F.~H.~L.}\ \bibnamefont
  {Koppens}}, \bibinfo {author} {\bibfnamefont {M.}~\bibnamefont
  {Pioro-Ladrière}}, \bibinfo {author} {\bibfnamefont {H.-P.}\ \bibnamefont
  {Tranitz}}, \bibinfo {author} {\bibfnamefont {I.~T.}\ \bibnamefont {Vink}},
  \bibinfo {author} {\bibfnamefont {S.}~\bibnamefont {Tarucha}}, \bibinfo
  {author} {\bibfnamefont {W.}~\bibnamefont {Wegscheider}}, \ and\ \bibinfo
  {author} {\bibfnamefont {L.~M.~K.}\ \bibnamefont {Vandersypen}},\ }\href
  {http://link.aps.org/doi/10.1103/PhysRevLett.101.226603} {\bibfield
  {journal} {\bibinfo  {journal} {Physical Review Letters}\ }\textbf {\bibinfo
  {volume} {101}},\ \bibinfo {pages} {226603} (\bibinfo {year}
  {2008})}\BibitemShut {NoStop}%
\bibitem [{\citenamefont {Liang}\ \emph {et~al.}(2011)\citenamefont {Liang},
  \citenamefont {Dong}, \citenamefont {Cheng}, \citenamefont {Gennser},
  \citenamefont {Cavanna},\ and\ \citenamefont {Jin}}]{Liang:2011}%
  \BibitemOpen
  \bibfield  {author} {\bibinfo {author} {\bibfnamefont {Y.~X.}\ \bibnamefont
  {Liang}}, \bibinfo {author} {\bibfnamefont {Q.}~\bibnamefont {Dong}},
  \bibinfo {author} {\bibfnamefont {M.~C.}\ \bibnamefont {Cheng}}, \bibinfo
  {author} {\bibfnamefont {U.}~\bibnamefont {Gennser}}, \bibinfo {author}
  {\bibfnamefont {A.}~\bibnamefont {Cavanna}}, \ and\ \bibinfo {author}
  {\bibfnamefont {Y.}~\bibnamefont {Jin}},\ }\href {\doibase
  doi:http://dx.doi.org/10.1063/1.3637054} {\bibfield  {journal} {\bibinfo
  {journal} {Applied Physics Letters}\ }\textbf {\bibinfo {volume} {99}},\
  (\bibinfo {year} {2011})}\BibitemShut {NoStop}%
\bibitem [{\citenamefont {Timp}\ \emph {et~al.}(1990)\citenamefont {Timp},
  \citenamefont {Behringer},\ and\ \citenamefont {Cunningham}}]{Timp:1990}%
  \BibitemOpen
  \bibfield  {author} {\bibinfo {author} {\bibfnamefont {G.}~\bibnamefont
  {Timp}}, \bibinfo {author} {\bibfnamefont {R.~E.}\ \bibnamefont {Behringer}},
  \ and\ \bibinfo {author} {\bibfnamefont {J.~E.}\ \bibnamefont {Cunningham}},\
  }\href {http://link.aps.org/doi/10.1103/PhysRevB.42.9259} {\bibfield
  {journal} {\bibinfo  {journal} {Physical Review B}\ }\textbf {\bibinfo
  {volume} {42}},\ \bibinfo {pages} {9259} (\bibinfo {year}
  {1990})}\BibitemShut {NoStop}%
\bibitem [{\citenamefont {Kurdak}\ \emph {et~al.}(1997)\citenamefont {Kurdak},
  \citenamefont {Chen}, \citenamefont {Tsui}, \citenamefont {Parihar},
  \citenamefont {Lyon},\ and\ \citenamefont {Weimann}}]{Kurdak:1997}%
  \BibitemOpen
  \bibfield  {author} {\bibinfo {author} {\bibfnamefont {C.}~\bibnamefont
  {Kurdak}}, \bibinfo {author} {\bibfnamefont {C.~J.}\ \bibnamefont {Chen}},
  \bibinfo {author} {\bibfnamefont {D.~C.}\ \bibnamefont {Tsui}}, \bibinfo
  {author} {\bibfnamefont {S.}~\bibnamefont {Parihar}}, \bibinfo {author}
  {\bibfnamefont {S.}~\bibnamefont {Lyon}}, \ and\ \bibinfo {author}
  {\bibfnamefont {G.~W.}\ \bibnamefont {Weimann}},\ }\href
  {http://link.aps.org/doi/10.1103/PhysRevB.56.9813} {\bibfield  {journal}
  {\bibinfo  {journal} {Physical Review B}\ }\textbf {\bibinfo {volume} {56}},\
  \bibinfo {pages} {9813} (\bibinfo {year} {1997})}\BibitemShut {NoStop}%
\bibitem [{\citenamefont {Birner}\ \emph {et~al.}(2007)\citenamefont {Birner},
  \citenamefont {Zibold}, \citenamefont {Andlauer}, \citenamefont {Kubis},
  \citenamefont {Sabathil}, \citenamefont {Trellakis},\ and\ \citenamefont
  {Vogl}}]{Birner:2007}%
  \BibitemOpen
  \bibfield  {author} {\bibinfo {author} {\bibfnamefont {S.}~\bibnamefont
  {Birner}}, \bibinfo {author} {\bibfnamefont {T.}~\bibnamefont {Zibold}},
  \bibinfo {author} {\bibfnamefont {T.}~\bibnamefont {Andlauer}}, \bibinfo
  {author} {\bibfnamefont {T.}~\bibnamefont {Kubis}}, \bibinfo {author}
  {\bibfnamefont {M.}~\bibnamefont {Sabathil}}, \bibinfo {author}
  {\bibfnamefont {A.}~\bibnamefont {Trellakis}}, \ and\ \bibinfo {author}
  {\bibfnamefont {P.}~\bibnamefont {Vogl}},\ }\href {\doibase
  10.1109/TED.2007.902871} {\bibfield  {journal} {\bibinfo  {journal} {IEEE
  Transactions on Electron Devices}\ }\textbf {\bibinfo {volume} {54}},\
  \bibinfo {pages} {2137} (\bibinfo {year} {2007})}\BibitemShut {NoStop}%
\bibitem [{\citenamefont {Schubert}\ and\ \citenamefont
  {Ploog}(1984)}]{Schubert:1984}%
  \BibitemOpen
  \bibfield  {author} {\bibinfo {author} {\bibfnamefont {E.~F.}\ \bibnamefont
  {Schubert}}\ and\ \bibinfo {author} {\bibfnamefont {K.}~\bibnamefont
  {Ploog}},\ }\href {http://link.aps.org/doi/10.1103/PhysRevB.30.7021}
  {\bibfield  {journal} {\bibinfo  {journal} {Physical Review B}\ }\textbf
  {\bibinfo {volume} {30}},\ \bibinfo {pages} {7021} (\bibinfo {year}
  {1984})}\BibitemShut {NoStop}%
\bibitem [{\citenamefont {Chadi}\ and\ \citenamefont
  {Chang}(1989)}]{Chadi:1989}%
  \BibitemOpen
  \bibfield  {author} {\bibinfo {author} {\bibfnamefont {D.~J.}\ \bibnamefont
  {Chadi}}\ and\ \bibinfo {author} {\bibfnamefont {K.~J.}\ \bibnamefont
  {Chang}},\ }\href {http://link.aps.org/doi/10.1103/PhysRevB.39.10063}
  {\bibfield  {journal} {\bibinfo  {journal} {Physical Review B}\ }\textbf
  {\bibinfo {volume} {39}},\ \bibinfo {pages} {10063} (\bibinfo {year}
  {1989})}\BibitemShut {NoStop}%
\bibitem [{\citenamefont {K.~Khachaturyan}\ and\ \citenamefont
  {Kaminska}(1989)}]{Khachaturyan:1989}%
  \BibitemOpen
  \bibfield  {author} {\bibinfo {author} {\bibfnamefont {E.~R.~W.}\
  \bibnamefont {K.~Khachaturyan}}\ and\ \bibinfo {author} {\bibfnamefont
  {M.}~\bibnamefont {Kaminska}},\ }in\ \href@noop {} {\emph {\bibinfo
  {booktitle} {Defects in Semiconductors}}},\ Vol.\ \bibinfo {volume} {38-41},\
  \bibinfo {editor} {edited by\ \bibinfo {editor} {\bibfnamefont
  {G.}~\bibnamefont {Ferenczi}}}\ (\bibinfo  {publisher} {Trans. Tech.
  Publications},\ \bibinfo {address} {Aedermannsdorf, Switzerland},\ \bibinfo
  {year} {1989})\ p.\ \bibinfo {pages} {1067}\BibitemShut {NoStop}%
\bibitem [{\citenamefont {Künzel}\ \emph {et~al.}(1983)\citenamefont
  {Künzel}, \citenamefont {Fischer}, \citenamefont {Knecht},\ and\
  \citenamefont {Ploog}}]{Kunzel:1983}%
  \BibitemOpen
  \bibfield  {author} {\bibinfo {author} {\bibfnamefont {H.}~\bibnamefont
  {Künzel}}, \bibinfo {author} {\bibfnamefont {A.}~\bibnamefont {Fischer}},
  \bibinfo {author} {\bibfnamefont {J.}~\bibnamefont {Knecht}}, \ and\ \bibinfo
  {author} {\bibfnamefont {K.}~\bibnamefont {Ploog}},\ }\href {\doibase
  10.1007/bf00617831} {\bibfield  {journal} {\bibinfo  {journal} {Applied
  Physics A}\ }\textbf {\bibinfo {volume} {32}},\ \bibinfo {pages} {69}
  (\bibinfo {year} {1983})}\BibitemShut {NoStop}%
\bibitem [{\citenamefont {Künzel}\ \emph {et~al.}(1984)\citenamefont
  {Künzel}, \citenamefont {Ploog}, \citenamefont {Wünstel},\ and\
  \citenamefont {Zhou}}]{Kunzel:1984}%
  \BibitemOpen
  \bibfield  {author} {\bibinfo {author} {\bibfnamefont {H.}~\bibnamefont
  {Künzel}}, \bibinfo {author} {\bibfnamefont {K.}~\bibnamefont {Ploog}},
  \bibinfo {author} {\bibfnamefont {K.}~\bibnamefont {Wünstel}}, \ and\
  \bibinfo {author} {\bibfnamefont {B.~L.}\ \bibnamefont {Zhou}},\ }\href
  {\doibase 10.1007/bf02656681} {\bibfield  {journal} {\bibinfo  {journal}
  {Journal of Electronic Materials}\ }\textbf {\bibinfo {volume} {13}},\
  \bibinfo {pages} {281} (\bibinfo {year} {1984})}\BibitemShut {NoStop}%
\bibitem [{\citenamefont {McWhorter}(1957)}]{McWhorter:1957}%
  \BibitemOpen
  \bibfield  {author} {\bibinfo {author} {\bibfnamefont {A.~L.}\ \bibnamefont
  {McWhorter}},\ }in\ \href@noop {} {\emph {\bibinfo {booktitle} {Semiconductor
  Surface Physics}}},\ \bibinfo {editor} {edited by\ \bibinfo {editor}
  {\bibfnamefont {R.~H.}\ \bibnamefont {Kingdton}}}\ (\bibinfo  {publisher}
  {University of Pennsylvania Press},\ \bibinfo {address} {Philadelphia, PA},\
  \bibinfo {year} {1957})\ pp.\ \bibinfo {pages} {207--228}\BibitemShut
  {NoStop}%
\bibitem [{\citenamefont {Jung}\ \emph {et~al.}(2004)\citenamefont {Jung},
  \citenamefont {Fujisawa}, \citenamefont {Hirayama},\ and\ \citenamefont
  {Jeong}}]{Jung:2004}%
  \BibitemOpen
  \bibfield  {author} {\bibinfo {author} {\bibfnamefont {S.~W.}\ \bibnamefont
  {Jung}}, \bibinfo {author} {\bibfnamefont {T.}~\bibnamefont {Fujisawa}},
  \bibinfo {author} {\bibfnamefont {Y.}~\bibnamefont {Hirayama}}, \ and\
  \bibinfo {author} {\bibfnamefont {Y.~H.}\ \bibnamefont {Jeong}},\ }\href
  {\doibase doi:http://dx.doi.org/10.1063/1.1777802} {\bibfield  {journal}
  {\bibinfo  {journal} {Applied Physics Letters}\ }\textbf {\bibinfo {volume}
  {85}},\ \bibinfo {pages} {768} (\bibinfo {year} {2004})}\BibitemShut
  {NoStop}%
\bibitem [{\citenamefont {Jia}\ \emph {et~al.}(1989)\citenamefont {Jia},
  \citenamefont {Li}, \citenamefont {Zhou}, \citenamefont {Gao}, \citenamefont
  {Kong}, \citenamefont {Yu},\ and\ \citenamefont {Chan}}]{Jia:1989}%
  \BibitemOpen
  \bibfield  {author} {\bibinfo {author} {\bibfnamefont {Y.~B.}\ \bibnamefont
  {Jia}}, \bibinfo {author} {\bibfnamefont {M.~F.}\ \bibnamefont {Li}},
  \bibinfo {author} {\bibfnamefont {J.}~\bibnamefont {Zhou}}, \bibinfo {author}
  {\bibfnamefont {J.~L.}\ \bibnamefont {Gao}}, \bibinfo {author} {\bibfnamefont
  {M.~Y.}\ \bibnamefont {Kong}}, \bibinfo {author} {\bibfnamefont {P.~Y.}\
  \bibnamefont {Yu}}, \ and\ \bibinfo {author} {\bibfnamefont {K.~T.}\
  \bibnamefont {Chan}},\ }\href {\doibase 10.1063/1.343672} {\bibfield
  {journal} {\bibinfo  {journal} {Journal of Applied Physics}\ }\textbf
  {\bibinfo {volume} {66}},\ \bibinfo {pages} {5632} (\bibinfo {year}
  {1989})}\BibitemShut {NoStop}%
\end{thebibliography}
\end{document}